\documentclass[twocolumn,floats,showpacs,preprintnumbers,amsmath,amssymb,aps]{revtex4}
\usepackage{epsfig,graphicx}
\usepackage{dcolumn}
\usepackage{bm}

\usepackage{epic,eepic,float}
\usepackage{amssymb,amsmath,multirow,rotate,color}

\usepackage{verbatim}

\begin{document}

\title{Influence of synaptic interaction on firing synchronization and
spike death in excitatory neuronal networks}

\author{Sheng-Jun Wang$^1$}
\author{Xin-Jian Xu$^2$}
\author{Zhi-Xi Wu$^3$}
\author{Zi-Gang Huang$^1$}
\author{Ying-Hai Wang$^1$}
\email{yhwang@lzu.edu.cn}

\affiliation{$^1$Institute of Theoretical Physics, Lanzhou
University, Lanzhou Gansu 730000, China\\
$^2$Department of Mathematics, College of Science, Shanghai University, Shanghai 200444, China\\
$^3$Department of Physics, Ume\aa\, University, 90187 Ume\aa,
Sweden}

\date{\today}

\begin{abstract}
We investigated the influence of efficacy of synaptic interaction
on firing synchronization in excitatory neuronal networks. We
found spike death phenomena, namely, the state of neurons transits
from limit cycle to fixed point or transient state. The phenomena
occur under the perturbation of excitatory synaptic interaction
that has a high efficacy. We showed that the decrease of synaptic
current results in spike death through depressing the feedback of
sodium ionic current. In the networks with spike death property
the degree of synchronization is lower and unsensitive to the
heterogeneity of neurons. The mechanism of the influence is that
the transition of neuron state disrupts the adjustment of the
rhythm of neuron oscillation and prevents further increase of
firing synchronization.
\end{abstract}


\maketitle

\section{introduction}

Synchronization of neural activity appears in different parts of
the mammalian cerebral cortex \cite{Singer}, and underlies
different neural processes in both normal and anomalous brain
functions \cite{Engel}. It has been suggested that synchronization
plays a vital role in information processing in the brain, e.g.,
processing information from different sensory systems to form a
coherent and unified perception of the external world
\cite{Singer, Engel, Malsburg, Roskies, Singer99, Rabinovich_RMP}.
On the other hand, synchronization has been detected in
pathological conditions such as Parkinson's disease \cite{Nini,
Brown07}. And epileptic seizures have long been considered
resulting from excessive synchronized brain activity \cite{Wong},
though some recent studies suggest that this picture may be an
over-simplification \cite{Netoff, Takeshita}. Therefore
understanding the mechanisms of synchronization may be a critical
step in elucidating how neural systems work \cite{Takeshita}. It
has stimulated a great deal of theoretical and numerical works,
such as the studies on the effects of the topological properties
of underlying networks \cite{Lago, Percha, Feldt, multinets} and
the dynamical properties of synaptic coupling \cite{Sakaguchi,
Zillmer}.

It was recently shown that the response time of synaptic couplings
influences the stability of synchronized oscillation in the
nonlocally coupled Hodgkin-Huxley (HH) equations \cite{Sakaguchi}.
If the response time of synaptic coupling is slower, synchronized
activity of the systems is instable for excitatory coupling.
However, the underlying dynamical mechanism of the influence is
not clear. In experimental studies \cite{Sayin}, it has been
suggested that the generation of prolonged epileptiform neuronal
synchronization is favored by lower efficacy of synaptic
transmission. The numerical studies \cite{Drongelen} in a detailed
computational model revealed that seizure-like activity occurs
when the excitatory synapses are weakened, and the results were
confirmed experimentally in mouse neocortical slices. According to
the common accepted assumption that synchronization of neuronal
activity underlies seizures, the dynamical mechanism of
synchronization may be useful for understanding the way the
biological neural system works.

In this work, we numerically investigated the dynamical mechanism
underlying the influence of synaptic efficacy on firing
synchronization in HH neuron networks. To do this, we first
studied the dynamics of the response of HH neuron to excitatory
synaptic current. When the efficacy of synapse is low, namely,
strength is weak and duration is short, the limit cycle is stable
to the perturbation of the synaptic current. When synaptic
efficacy is high, synaptic current can induce the transition of
the neurons from limit cycle to fixed point or transient state.
The transition is determined by dynamics of neuron's ionic
channel. The decrease of synaptic current depresses the feedback
of sodium ionic current which is responsible for the initiation of
the spike. For simplicity we will refer to the transitions as
spike death.

In neuronal networks, synaptic input of a neuron is the
accumulation of the currents received from all presynaptic
neurons. When the coherence of firing time of neurons is enhanced
by the excitatory interaction, the synaptic input of neurons
transforms from the fluctuating waveform into the pulse shape like
the signal produced by one synapse. If synaptic efficacy is high,
the input signal can induce spike death of the neuron. Then spike
death disorders the adjustment of the rhythm of neurons and
prevents neurons from firing spikes synchronously. In contrast,
for synapses of lower efficacy, the duration of synaptic current
is too short to induce spike death of neurons. Additionally, the
firing synchronization is different from synchronous activity of
oscillators for the existence of the transitions of neuron's
state.

The paper is organized as follows. The HH neuron model and the
synaptic coupling were introduced in Sec. II. The response of a HH
neuron to synaptic current was investigated in Sec. III. The
influence of the dynamics of neurons on firing synchrony was shown
in Sec. IV. Discussion and conclusion were given in Sec. V.

\section{model}
To investigate the dynamics of a neuron under the perturbation of
synaptic stimulus, we adopted a system consisting of a HH neuron
and a synapse. The HH neuron was originally proposed for the giant
axon in a squid \cite{HH}. It serves as a paradigm for the spiking
neuron models that based on nonlinear conductances of ion
channels. The model describes the evolution of the membrane
potential $V(t)$ and can be written as
\begin{eqnarray}
C \frac{dV}{dt}=I_{ion}+I_{stim}+I_{syn},
\end{eqnarray}
where $I_{ion}$ is the ionic current, $I_{stim}$ is the external
current and $I_{syn}$ is the synaptic current. The ionic current
describes the ion channel on the membrane and is defined as
\begin{equation}
I_{ion}= -g_{Na}m^3h(V-E_{Na})-g_Kn^4(V-E_K)-g_{l}(V-E_{l}),
\end{equation}
where $g_{Na}$, $g_{K}$ and $g_{l}$ are the maximum conductances
for the sodium, potassium and leak currents, $E_{Na}$, $E_{K}$ and
$E_{l}$ are the corresponding reversal potentials. $m$ and $h$ are
the activation and inactivation variables of the sodium current
and $n$ is the activation variable of the potassium current. The
gating variables $y=$ $m$, $h$, $n$ satisfy the differential
equation
\begin{equation}
\frac{dy(t)}{dt}=\alpha_y[1-y(t)]-\beta_yy(t), \label{eq:dy}
\end{equation}
with nonlinear functions $\alpha_y$ and $\beta_y$ given by
\begin{eqnarray}
\alpha_m &=& 0.1(V+40)/\{1-\exp[-(V+40)/10]\}, \\
\beta_m &=& 4\exp[-(V+65)/18], \\
\alpha_h &=& 0.07\exp[-(V+65)/20], \\
\beta_h &=& 1/\{1+\exp[-(V+35)/10]\}, \\
\alpha_n &=& 0.01(V+55)/\{1-\exp[-(V+55)/10] \}, \\
\beta_n &=& 0.125\exp[-(V+65)/80].
\end{eqnarray}
The parameter values are $E_{Na}=50$ mV, $E_K=-77$ mV,
$E_{l}=-54.4$ mV, $g_{Na}=120$ mS/cm$^2$, $g_K=36$ mS/cm$^2$,
$g_{l}=0.3$ mS/cm$^2$, and $C=1$ $\mu$F/cm$^{2}$ \cite{Gerstner,
Hansel}.

The external current $I_{stim}$ determines the firing rate of the
neuron. In the absence of synaptic coupling $I_{syn}$, the HH
neuron has the following bifurcation diagram as a function of
$I_{stim}$: In the parameter regions $I_{stim}<I_0$ and
$I_{stim}>I_2$ fixed point is the global attractor. For
$I_0<I_{stim}<I_1$ the neuron possesses coexisting stable
attractors, the fixed point and the limit cycle which are
separated by an unstable limit cycle. For $I_1<I_{stim}<I_2$ fixed
point becomes unstable. The values of bifurcation points are
$I_0\approx 6.2$ $\mu$A/cm$^2$, $I_1 \approx 9.8$ $\mu$A/cm$^2$
and $I_2\approx154$ $\mu$A/cm$^2$ \cite{Hassard}.

We adopted the synaptic current $I_{syn}$ described by an alpha
function \cite{Hansel}. The alpha function synapse is a
phenomenological model based on an approximate correspondence of
the time course of the waveform to physiological recordings of the
postsynaptic response \cite{Destexhe}. The equation of the synapse
is like:
\begin{equation}
I_{syn}(t)=-g_{syn}\alpha(t-t_{in})(V(t)-E_{syn}),
\end{equation}
with
\begin{equation}\label{alpha}
\alpha(t)=(t/\tau)\exp(-t/\tau)\Theta(t).
\end{equation}
Where $\tau$ is the characteristic time of the interaction,
$\Theta(t)$ is the Heaviside step function, and $t_{in}$ is the
beginning time of the synaptic interaction, i.e., the firing time
of the presynaptic neuron (all delays are neglected). The synaptic
effect is traditionally classified as excitatory or inhibitory
depending on the value of $E_{syn}$. Here, we took $E_{syn}=30$ mV
for excitatory synapses, and $-80$ mV for inhibitory ones. Eq.
(\ref{alpha}) yields pulses with the maximum value of $e^{-1}$ at
$t=t_{in}+\tau$ and with the half-width of 2.45$\tau$
\cite{Hasegawa}. So $\tau$ characterizes the duration of the
synaptic interaction. For the alpha function synapse, we equated
the synaptic efficacy to the maximum synaptic conductances
$g_{syn}$ and the characteristic time $\tau$. The high efficacy
means that synaptic current possesses strong strength and long
duration.

\section{spike death of neuron}

\begin{figure}
\includegraphics[width=\columnwidth]{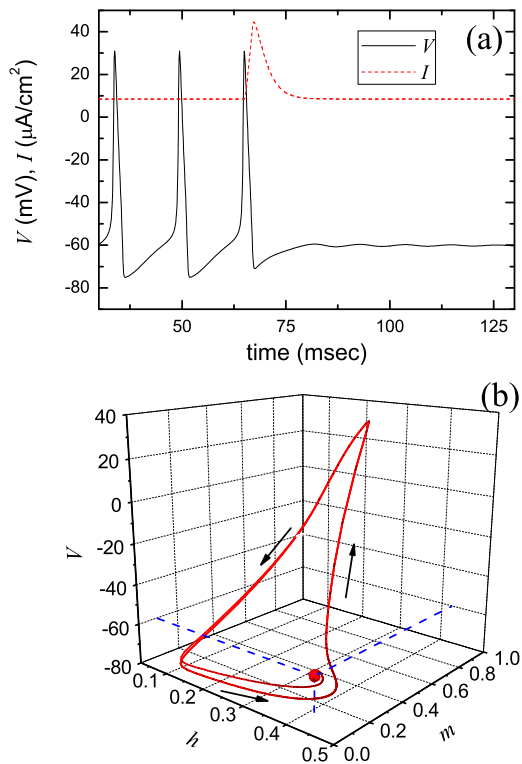}
\caption{\label{fig1}(Color online) (a) The response of bistable
HH neuron to excitatory synaptic current. The symbol $I$ denotes
the sum of the external current $I_{stim}$ and the synaptic
current $I_{syn}$. (b) The corresponding phase portrait. The
values of parameters are $\tau=2$ msec, $g_{syn}=1$ mS/cm$^2$ and
$I_{stim}=8.5$ $\mu$A/cm$^2$.}
\end{figure}

We focused on the dynamics of the system in the parameter region
near the bifurcation point $I_1$. Firstly, we studied the response
of bistable neuron ($I_0 < I_{stim}< I_1$) to excitatory synaptic
current. In simulations, firing was identified as membrane
potential $V$ is over 20 mV. When the neuron fired a spike, we
triggered a pulse of synaptic current into it. We observed two
types of dynamics of the response, which depended on the efficacy
of synapse. For slow response time and strong synaptic strength,
the neuron transited from the limit cycle to the fixed point. The
transition of neuron's state is shown in Figs. \ref{fig1}. In Fig.
\ref{fig1}(a) the response of the neuron to the synaptic current
is represented by the membrane potential $V$. In the figure, the
synaptic current is illustrated by the pulse added on the external
current. One can see that the periodic firing eliminated after the
synaptic current was injected. The value of membrane potential
tended to -61.15 mV through subthreshold oscillation. In Fig.
\ref{fig1}(b) the transition of neuron's state is shown in a
three-dimensional space ($V$, $h$, $m$) which is a projection of
the phase space ($V$, $h$, $m$, $n$). The trajectory left the
limit cycle and was attracted into the basin of the fixed point.
At last, through transient process, the trajectory stopped at the
fixed point which is indicated by the dashed lines in the figure.
The fixed point was $(V,h,m,n)$=(-60.15, 0.423, 0.092, 0.394) as
the parameter $I_{stim}=8.5$ $\mu$A/cm$^2$. On the other hand, for
systems with quick response time and weak synaptic strength, the
transition did not occur and the trajectory was attracted back the
limit cycle from weak perturbation.

\begin{figure}
\includegraphics[width=\columnwidth]{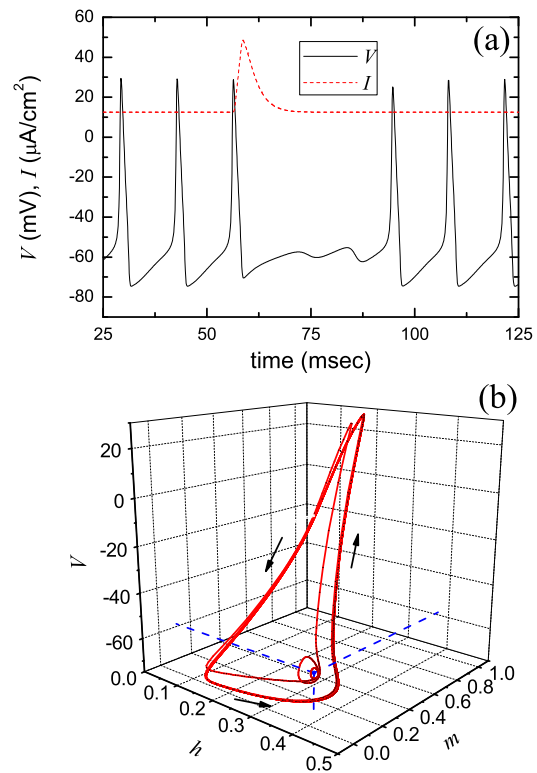}
\caption{\label{fig2} (Color online) (a) The response of the
neuron with $I_{stim}=12.5$ $\mu$A/cm$^2$ to excitatory synaptic
current. (b) The corresponding phase portrait. The synaptic
parameters are the same as Fig. \ref{fig1}.}
\end{figure}

When the external signal $I_{stim}$ is larger than but near $I_1$,
the neuron possesses a stable limit cycle and an unstable fixed
point. In this case there were also two types of dynamics of
response. For high synaptic efficacy, the neuron exhibited a
transient behavior when it received the synaptic current. Fig.
\ref{fig2} (a) shows that the membrane potential responded to the
synaptic current with a transient subthreshold oscillation. The
subthreshold oscillation interrupted the periodic firing of the
neuron. In phase space the transient behavior was a motion around
the unstable fixed point. This is explicitly shown in the
corresponding 3D phase space ($V$, $h$, $m$). In Fig.
\ref{fig2}(b) one can see that the trajectory of the neuron left
the limit cycle and transiently moved around the unstable fixed
point which is indicated by the dashed lines. The unstable fixed
point was $(V,h,m,n)$=(-58.704, 0.374, 0.108, 0.417) as the
parameter $I_{stim}=12.5$ $\mu$A/cm$^2$. On the other hand, for
low synaptic efficacy, the coupling cannot induce the transient
motion around the unstable fixed point. Like bistable neurons, the
trajectory returned to limit cycle from weak perturbation.

\begin{figure}
\includegraphics[width=\columnwidth]{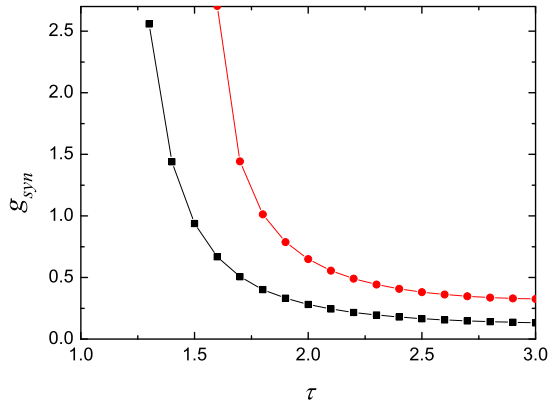}
\caption{\label{fig3} (Color online) The dynamics diagram of
minimal synaptic intensity $g_{syn}$ vs the characteristic time
$\tau$ for the neurons with the external current $I_{stim}=8.5$
$\mu$A/cm$^2$ (squares) or $I_{stim}=12.5$ $\mu$A/cm$^2$
(circles).}
\end{figure}

Fig. \ref{fig3} shows the boundary between the two types of
dynamics on the parameter plane $g_{syn}$ vs $\tau$. For the
bistable neuron of the external current $I_{stim}=8.5$
$\mu$A/cm$^2$, the boundary between the two types of dynamics is
represented by squares. Above the curve, neurons responded to
synaptic currents with transitions between attractors. Inversely,
the limit cycle of neurons was stable to the perturbation. For the
neuron of the external current $I_{stime}=12.5$ $\mu$A/cm$^2$, the
boundary is shown by circles. It is notable that stronger strength
and longer duration of synaptic current were needed by the
transitions of neuron's state. Based on numerical simulations, we
obtained that the transitions of neuron's state can be induced by
synaptic current in the parameter region $6.2$ $\mu$A/cm$^2$
$<I_{stim}<28.8$ $\mu$A/cm$^2$. Above the upper boundary of the
region, the transient motion around the unstable fixed point
cannot be induced by the synaptic current.

In the following, we referred to the phenomena shown in both Fig.
\ref{fig1} and Fig. \ref{fig2} as spike death. Different from
oscillator death which is the quenching of oscillation of coupled
systems, spike death is the behavior of a single neuron. And spike
death includes both the transition between stable attractors and
the transient behavior.

\begin{figure}
\includegraphics[width=\columnwidth]{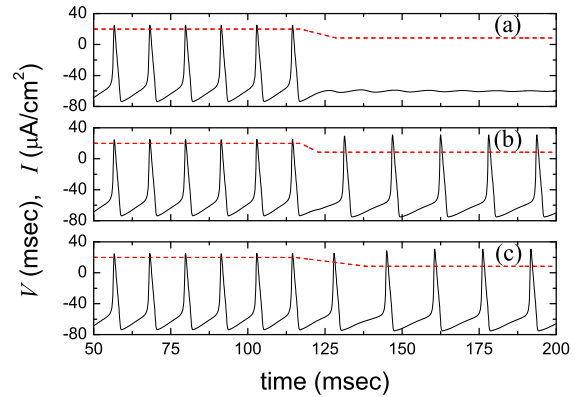}
\caption{\label{fig4} (Color online) The change of the activity of
HH neuron as external current decreases. The external current is
reduced from 20 $\mu$A/cm$^2$ to 8.5 $\mu$A/cm$^2$ with the slop
(a) -1 $\mu$A/(cm$^2$ msec), (b) -2 $\mu$A/(cm$^2$ msec), or (c)
-0.5 $\mu$A/(cm$^2$ msec).}
\end{figure}

Next we gave a qualitative interpretation of the spike death
phenomena. Although the synaptic current consists of a rise and a
decay stage, we want to show that the decrease of the synaptic
current induces the transition of neuronal activity. We
illustrated the role of the decreased current in Fig. \ref{fig4}.
In the simulations, external currents linearly decreased from 20
$\mu$A/cm$^2$ to 8.5 $\mu$A/cm$^2$, and synaptic current was
absent. The time of beginning to decrease and the different slopes
of the decrease current were chosen to represent the distinct
duration and rate of decay of the current. In Fig. \ref{fig4} (a),
the activity of the neuron transited from the periodic firing to
silent state through transient subthreshold oscillation. The
decrease of current induced the spike death of the neuron. In Fig.
\ref{fig4} (b), the decay of the current occurred in the
refractory period of the neuron. The decreased current did not
depress the next spike. This is similar to the scenario of the
synaptic current of short duration. So spike death requires that
the decrease of current occurs at the end of refractory period and
the stage of the initiation of next spike. In Fig. \ref{fig4} (c),
the external current decreased with the slope -0.5 $\mu$A/(cm$^2$
msec). Comparing with Fig. \ref{fig4} (a), the rate of decrease
was small. Although the current decreased at the stage of
initiating a spike, it did not depress the firing of the neuron.
Thus spike death requires that the rate of decrease of current is
large at the stage of initiating spikes.

To interpret the effect of the decrease of current on the
oscillation of neurons, we reviewed the generation of spikes. For
a fixed value of membrane potential $V$, the variable $y$ (= $m$,
$h$, $n$) approaches the value
$y_{0}(V)=\alpha_{y}(V)/[\alpha_{y}(V)+\beta_{y}(V)]$ with the
time constant $\tau_{y}(V)=[\alpha_{y}(V)+\beta_{y}(V)]^{-1}$. The
variable $m_0(V)$ increases with $V$, and the corresponding time
scale $\tau_{m}$ is littler than $\tau_h$ and $\tau_n$. If
external current injects into the cell and raises the membrane
potential $V$, the conductance of sodium channels $g_{Na}m^{3}h$
increases due to increasing $m$. Then sodium ions flow into the
cell and raise membrane potential even further. If this positive
feedback is large enough, a spike is initiated \cite{Gerstner}.
When external current decreases at the onset of the generation of
a spike, the raise of membrane potential slows down. Then the
increase of $m$ is slowed down. So the positive feedback is
weakened. If the current decreases quickly, the spike of neuron
can be depressed. Therefore, for depressing the positive feedback
of sodium ionic current, the decrease of current must occur at the
onset of generation of spike, and the current must decrease in a
large rate. This interpretation is consistent with above simulated
results that the strong strength and the long duration of synaptic
current are necessary for spike death.

\section{influence on firing synchronization}

Next we investigated the influence of spike death on the firing
synchronization in neuronal networks. We considered a directed
random network consisting of $N$ nonidentical HH neurons. The
network was generated as following: With a probability $p$, we
connected each of the probable directed couplings (such as the one
from the $j$th neuron to the $i$th neuron). The network is
described by adjacency matrix $\{a^{ij}\}$ which entry $a^{ij}$ is
equal to 1 when the coupling from $j$ to $i$ exists, and zero
otherwise. In the directed network, $a^{ij}$ can be not equal to
$a^{ji}$, and the signal travels in only the direction from $j$ to
$i$ if $a^{ij}$=1 and $a^{ji}$=0. The signal received by the
neuron $i$ is the accumulation of all input synaptic currents,
which is defined as
\begin{equation}
I_{syn}^i(t)=-\frac{1}{q^i}\sum_{j=1}^{N} a^{ij} g_{syn}
\alpha(t-t_{in}^j) (V^i(t)-E_{syn}^{ij}),
\end{equation}
where $q^i$ is the rescaled factor which equals the number of
inputting synapses of neuron $i$ ($i=1,\ldots,N$). $t_{in}^j$ is
the latest firing time of presynaptic neuron $j$. $E_{syn}^{ij}$
is the reverse potential of the synapse connecting $j$ to $i$.

To study the global behavior of neuronal networks we computed the
average activity $\overline{V}(t)=(1/N)\sum_{i=1}^N V^i(t)$ of the
network. The amplitude of average activity can intuitionally
reveal the coherence of activity of neurons, which is defined as
\cite{Lago}
\begin{equation}
\sigma^2=\frac{1}{T_2-T_1}\int_{T_1}^{T_2}[\langle
\overline{V}(t)\rangle_t - \overline{V}(t)]^2 dt,
\end{equation}
where the angle brackets denote temporal average over the
integration interval. In this work we studied the firing synchrony
for the reason that neuronal states may transit away from the
limit cycle and the synchronization of oscillators will be
disrupted. Here we focused on the coherence of firing time of
neurons. We adopted the average cross correlation of firing time
of neurons \cite{XJWang,SWang} to quantify the degree of firing
synchronization. Average cross correlation is obtained by
averaging the pair coherence $K_{ij}(\gamma)$ between neuron $i$
and $j$, i.e.,
\begin{equation}
K=\frac{1}{N(N-1)} \sum_{i=1}^{N} \sum_{j=1, j\neq i}^{N}
K_{ij}(\gamma).
\end{equation}
The pair coherence $K_{ij}(\gamma)$ is defined as
\begin{equation}
K_{ij}(\gamma)=\frac{\sum_{l=1}^{k}X(l)Y(l)}{[\sum_{l=1}^{k}X(l)\sum_{l=1}^{k}Y(l)]^{\frac{1}{2}}},
\end{equation}
which is measured by the cross correlation of spike trains at zero
time lag within a time bin $\gamma$. To transform the neuronal
activity into spike train, the interval $T_2-T_1$ is divided into
$k$ bins of $\gamma=1$ msec. Then spike trains of neurons $i$ and
$j$ are given by $X(l)=0$ or 1 and $Y(l)=0$ or 1 ($l=1,\ldots,
k$), where 1 represents a spike generates in the bin and 0
otherwise.

\begin{figure}
\includegraphics[width=\columnwidth]{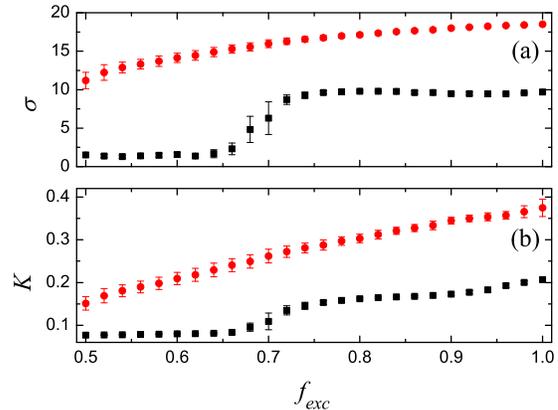}
\caption{\label{fig5} (color online) (a) The amplitude of average
activity $\sigma$ versus the fraction of excitatory neuron
$f_{exc}$. Squares (cycles) represented the relations in networks
with $\tau$=2 msec (1 msec). The error bars were the standard
deviation across 20 realizations. (b) The relation between average
cross correlation $K$ and the fraction of excitatory neuron
$f_{exc}$.}
\end{figure}

We numerically investigated the collective activity of networks
versus the fraction of excitatory neurons $f_{exc}$ in the
networks. In general, excitatory synapses tend to synchronize the
activity of neurons in networks. However, it was shown that in
neural systems different types of dynamics of response of neurons
to couplings produce different synchrony property \cite{Hansel95}.
Here we studied the influence of the new type of response, spike
death, on the formation of synchrony as the excitatory
interactions increase. In simulations, the networks consisted of
1000 neurons and connecting probability was $p$=0.01. The synaptic
conductance took the value $g_{syn}$=1 mS/cm$^2$. As an example,
we used the synapses with the characteristic time $\tau$=2 msec or
$\tau$=1 msec to generate neuronal networks with or without spike
death, respectively. To ensure that the neurons had nonidentical
properties, the external currents $I_{stim}^i$ were in $(8.0,
12.0)$ $\mu$A/cm$^2$ and were generated at random.

In Fig. \ref{fig5}(a) we plotted the amplitude of average activity
$\sigma$ versus the fraction $f_{exc}$ of excitatory neurons.
Squares represents the results obtained in networks with $\tau$=2
msec, and circles represents the results with $\tau$=1 msec. In
Fig. \ref{fig5}(b) we did the same for average cross correlation
$K$. The relations showed that the coherence of activity increases
with the fraction of excitatory neurons. It is notable that the
networks with $\tau$=2 msec, i.e., with the spike death property,
had obviously lower values of $\sigma$ and $K$ than networks with
$\tau=1$ msec. So the degree of synchrony in networks with spike
death property was obviously lower than networks without the
property.

\begin{figure}
\includegraphics[width=\columnwidth]{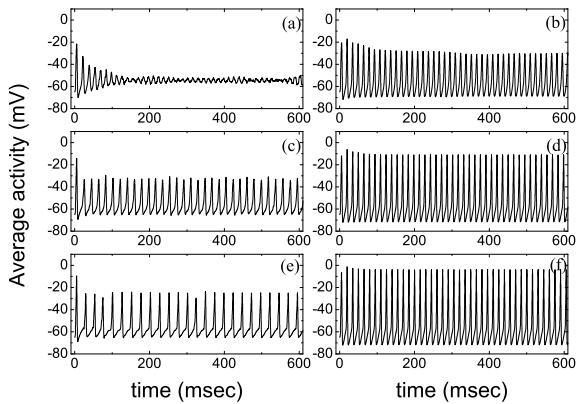}
\caption{\label{fig6} Average membrane potential of the networks
with $\tau=2$ msec (left) and $\tau=1$ msec (right). The fraction
of excitatory neurons is $f_{exc}=0.5$, 0.8, 1.0 from top to
bottom.}
\end{figure}

The difference of the degree of synchrony between two types of
networks is shown more intuitionally by the average activity in
Fig. \ref{fig6}. We made neurons in the networks beginning to
oscillate with a high degree of coherence. In simulations the
first firing time of neurons randomly distributed in (0, 5) msec.
When $f_{exc}$=0.5, the average activity of the network with
$\tau=2$ msec decreased obviously and the network changed quickly
to oscillating randomly, as shown in Fig. \ref{fig6}(a). In the
network without spike death property, coherent oscillation
\cite{Lago} appeared, as shown in Fig. \ref{fig6}(b). When the
fraction of excitatory neurons increased, the excitatory
interaction enhanced the average activity in the networks both
with and without the spike death property, while the average
activity in the latter was obviously larger. Furthermore, the
frequency of the average activity $\overline{V}$ in networks with
$\tau=1$msec was higher than the frequency in networks with
$\tau$=2 msec.

\begin{figure}
\includegraphics[width=\columnwidth]{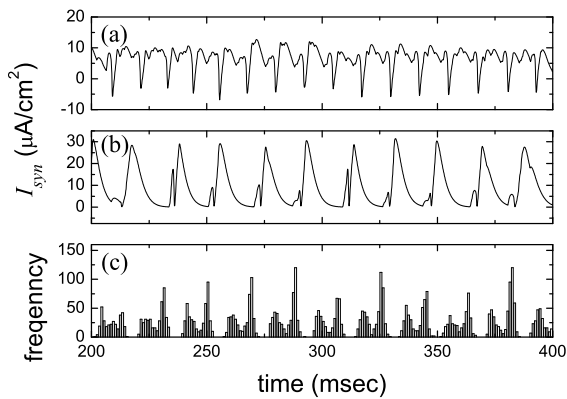}
\caption{\label{fig7} (a) The input synaptic current of a randomly
chosen neuron from the network with $\tau=2$ msec, $f_{exc}=0.5$.
(b) The input synapse current when $f_{exc}$ increased to 0.8. (c)
The histogram of spike death events in the network with
$\tau=2$msec and $f_{exc}=0.8$.}
\end{figure}

The synchrony properties can be qualitatively explained by spike
death. In networks, if the fraction of excitatory neuron is
increased, the coherence of firing time of neurons is enhanced.
Then the input synaptic current of a neuron may transform from
fluctuating waveform into a smooth pulse. If the pulse of the
accumulated synaptic current has long enough duration to depress
the next spike, the synaptic current may lead to spike death.
Therefore the synaptic current disorders the adjustment of the
rhythm of neurons' firing and prevents synchronization. To
demonstrate this, we showed the synaptic currents and spike death
events in Fig. 7. In Fig. 7(a) and 7(b) we plotted the input
synaptic currents of a neuron, which was randomly chosen in
networks with $f_{exc}$=0.5 and $f_{exc}$=0.8 respectively. The
transform from fluctuate waveform to smooth pulse was observed. In
simulations we took the peak of subthreshold oscillation of a
neuron as a spike death event. Fig. 7(c) shows the histogram of
frequency of spike death events, i.e., the number of spike death
events in 1 msec. The spike death events periodically appeared
with the same rhythm as the average activity of the network. In
contrast, spike death events cannot be obtained in the networks of
low synaptic efficacy. Thus the spike death can explain the above
synchrony property.

If network consists of identical neurons, the influence of spike
death also exists and is more remarkable. Fig. 8(a) shows the
relation of $K$ to $f_{exc}$ of the networks in which the external
currents of all neurons was $I_{stim}$=10.0$\mu$A/cm$^2$. One can
see that the values of $K$ of the networks with $\tau$=2 msec
(squares) was similar to Fig. 5. For networks with $\tau$=1 msec
(circles), however, the value of $K$ tended to 1 as $f_{exc}$
increases. We calculated the degree of synchrony as a function of
the heterogeneity of neurons in purely excitatory networks. In
simulations the value of external currents $I_{stim}$ distributed
in the region $(10.0-0.5w, 10.0+0.5w)$ $\mu$A/cm$^2$. Fig. 8(b)
shows the relation between $K$ and the width $w$ of the parameter
region of $I_{stim}$. For networks with $\tau=2$msec (squares),
the degree of synchrony was unsensitive to the heterogeneity of
neurons. In contrast, the degree of synchrony in networks with
$\tau=1$msec (circles) remarkably increased and tended to 1 as the
heterogeneity of neurons decreased. For networks with
$\tau=1$msec, the relation between $K$ and the width $w$ of the
parameter region was fitted to the first order exponential decay
curve. The fitted curve (dashed line) is $K=Aexp(-w/B)+K_0$ with
$K_0$ = 0.362 $\pm$ 0.005, $A$ = 0.595 $\pm$ 0.007 and $B$ = 1.017
$\pm$ 0.030. The remarkable difference between two kinds of
networks shows that spike death can effectively prevent the firing
synchronization in neuronal networks.

\begin{figure}
\includegraphics[width=\columnwidth]{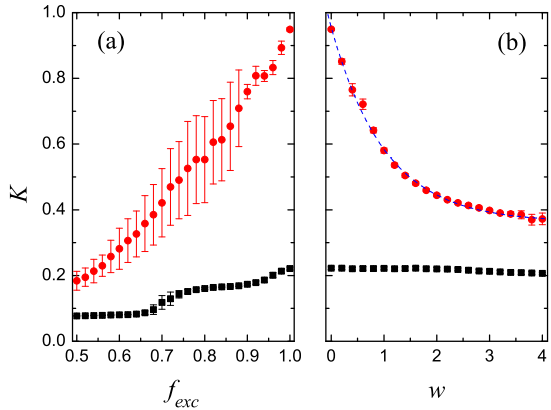}
\caption{\label{fig8} (Color online) (a): The average cross
correlation $K$ versus the fraction of excitatory neurons for the
networks consisting of identical neurons. Squares (circles)
represent the relation in networks with $\tau=2$msec (1msec). The
error bars were the standard deviation across 20 realizations.
(b): The average cross correlation $K$ changes with the width $w$
of parameter region of $I_{stim}$. Dashed line is the fitted
exponential curve.}
\end{figure}

\section{discussion and conclusion}

We have used HH neuron networks to investigate the dynamical
origin of the influence of synaptic efficacy on the firing
synchronization. A new dynamics of response of neurons to
coupling, spike death, was suggested as a possible mechanism
underlying the influence. When the firing time of neurons is so
coherent that synaptic currents have a pulse waveform in
excitatory networks, synaptic current induces the transition of
neuron's state from limit cycle to fixed point or transient state.
The transitions disrupt the adjustment of rhythm of neurons'
oscillations and prevent further increase of firing
synchronization.

We studied the dynamics of the HH neuron responding to the
excitatory synaptic perturbation. We numerically demonstrated that
the synapse of high efficacy, i.e., large characteristic time and
strong strength, induces spike death of the neuron. For bistable
neurons, spike death means the transition from limit cycle to
fixed point. For the neuron with unstable fixed point, spike death
means the transition from limit cycle to transient state. The
transient state is the motion around the unstable fixed point in
phase space. Spike death of neurons results from the decrease of
synaptic current, which depresses the feedback of sodium ionic
current at the stage of initiating a spike.

We demonstrated the influence of spike death on the degree of
firing synchronization. In simulations we considered the networks
with or without the spike death property, which were generated
using synapses of characteristic time $\tau$=2 msec and $\tau$=1
msec, respectively. The degree of synchrony of the former was
lower. This is consistent with results of \cite{Sakaguchi} that
synchronous state is not stable for the excitation coupling of
slow response time. However, we also showed that for slow response
time, the degree of synchrony increased with the fraction of
excitatory, and oscillation rate of whole network slowed down. Our
main results are that in the network with $\tau$=2 msec, the spike
death events were observed. And spike death can explain the
mechanism of preventing the raise of the degree of synchrony.
Related synchrony properties was found also in weakly coupled HH
neurons \cite{Hansel95, Hansel}. In the case of weak coupling, the
phase of neuron was perturbed by couplings, but the oscillation of
neuron was not destroyed. In contrast, the new dynamical mechanism
we suggested is proper for strong coupling and underlies the
synchrony of the interrupted oscillations. Additionally, it is
notable that, for the existence of spike death, the firing
synchronization of neuronal networks is different from the usual
oscillator synchronization in which each oscillator is stable to
perturbations \cite{synchrony}.

Our work relates to that of Drover et al. \cite{Drover}. In the
elegant work, using a simplified neuron model of two variables,
they proposed a mechanism for slowing firing down. They found that
slow decay synaptic variable induces that trajectory is attracted
toward the unstable fixed point of the simplified model. This is
similar with the transient behavior of HH neuron we proposed here.
However, with their mechanism, synaptic excitation is strongly
synchronizing in networks in contrast with that spike death
prevents synchrony. In neural networks, collective behaviors
sensitively depend on the intrinsic dynamics of neurons
\cite{Drover}, and many types of response of neuron to synaptic
coupling may exist \cite{Hansel95}. It is interesting to make
further studies on the relations among different responses and
their effects on the collective behaviors of networks.

The variability of strength of synapses was not involved in the
present investigation. The synaptic strength is often affected by
the activity of neurons through synaptic plasticity
\cite{plasticity} and synaptic adaption \cite{adaption}. The
effect of changes of synaptic strength will be studied elsewhere.

As mentioned above, the phenomena that low efficacy of synapses
favors the generation of neuronal synchronization underlying
seizure were obtained in experiments and numerical simulations
\cite{Sayin, Drongelen}. Here we equated the strength and
especially the characteristic time of synapses to the synaptic
efficacy, and studied the mechanism by which synapses influence
firing synchronization. The mechanism of the influence may have
potentially values for understanding the way the realistic neural
system works.

\begin{acknowledgements}
We thank L. Wang for many helpful discussions and C.-F. Feng for
comment reading. This work was supported by the NSF of China,
Grant No. 10775060. X.-J.X acknowledges financial support from FCT
(Portugal), Grant No. SFRH/BPD/30425/2006.
\end{acknowledgements}

\end{document}